\documentclass[aps,prl,showpacs,showkeys,amsmath,amssymb,
twocolumn,
floatfix,
superscriptaddress
]{revtex4-1}
\usepackage{graphicx}
\usepackage{times}
\usepackage{verbatim}
\usepackage{color}
\usepackage{cancel}
\usepackage[normalem]{ulem}
\begin{document}


\title{Measurement of the Reactor Antineutrino Flux and Spectrum at Daya Bay}

%%% Generation information for this file
%%% Mon Aug 17 00:00:00 2015
%%% /Users/djaffe/Documents/Reactor/work/trunk/NuWa-trunk/dybgaudi/Documentation/AuthorList/Reactor2015PRL
%%% ../AuthGen.py dybCollaborationList_20140301.xls 20140301 
\newcommand{\ECUST}{\affiliation{Institute of Modern Physics, East China University of Science and Technology, Shanghai}}
\newcommand{\IHEP}{\affiliation{Institute~of~High~Energy~Physics, Beijing}}
\newcommand{\Wisconsin}{\affiliation{University~of~Wisconsin, Madison, Wisconsin, USA}}
\newcommand{\Yale}{\affiliation{Department~of~Physics, Yale~University, New~Haven, Connecticut, USA}}
\newcommand{\BNL}{\affiliation{Brookhaven~National~Laboratory, Upton, New York, USA}}
\newcommand{\NTU}{\affiliation{Department of Physics, National~Taiwan~University, Taipei}}
\newcommand{\NUU}{\affiliation{National~United~University, Miao-Li}}
\newcommand{\Dubna}{\affiliation{Joint~Institute~for~Nuclear~Research, Dubna, Moscow~Region}}
\newcommand{\CalTech}{\affiliation{California~Institute~of~Technology, Pasadena, California, USA}}
\newcommand{\CUHK}{\affiliation{Chinese~University~of~Hong~Kong, Hong~Kong}}
\newcommand{\NCTU}{\affiliation{Institute~of~Physics, National~Chiao-Tung~University, Hsinchu}}
\newcommand{\NJU}{\affiliation{Nanjing~University, Nanjing}}
\newcommand{\TsingHua}{\affiliation{Department~of~Engineering~Physics, Tsinghua~University, Beijing}}
\newcommand{\SZU}{\affiliation{Shenzhen~University, Shenzhen}}
\newcommand{\NCEPU}{\affiliation{North~China~Electric~Power~University, Beijing}}
\newcommand{\Siena}{\affiliation{Siena~College, Loudonville, New York, USA}}
\newcommand{\IIT}{\affiliation{Department of Physics, Illinois~Institute~of~Technology, Chicago, Illinois, USA}}
\newcommand{\LBNL}{\affiliation{Lawrence~Berkeley~National~Laboratory, Berkeley, California, USA}}
\newcommand{\UIUC}{\affiliation{Department of Physics, University~of~Illinois~at~Urbana-Champaign, Urbana, Illinois, USA}}
\newcommand{\RPI}{\affiliation{Department~of~Physics, Applied~Physics, and~Astronomy, Rensselaer~Polytechnic~Institute, Troy, New~York, USA}}
\newcommand{\SJTU}{\affiliation{Shanghai~Jiao~Tong~University, Shanghai}}
\newcommand{\BNU}{\affiliation{Beijing~Normal~University, Beijing}}
\newcommand{\WM}{\affiliation{College~of~William~and~Mary, Williamsburg, Virginia, USA}}
\newcommand{\Princeton}{\affiliation{Joseph Henry Laboratories, Princeton~University, Princeton, New~Jersey, USA}}
\newcommand{\VirginiaTech}{\affiliation{Center for Neutrino Physics, Virginia~Tech, Blacksburg, Virginia, USA}}
\newcommand{\CIAE}{\affiliation{China~Institute~of~Atomic~Energy, Beijing}}
\newcommand{\SDU}{\affiliation{Shandong~University, Jinan}}
\newcommand{\NanKai}{\affiliation{School of Physics, Nankai~University, Tianjin}}
\newcommand{\UC}{\affiliation{Department of Physics, University~of~Cincinnati, Cincinnati, Ohio, USA}}
\newcommand{\DGUT}{\affiliation{Dongguan~University~of~Technology, Dongguan}}
\newcommand{\XJTU}{\affiliation{Xi'an Jiaotong University, Xi'an}}
\newcommand{\UCB}{\affiliation{Department of Physics, University~of~California, Berkeley, California, USA}}
\newcommand{\HKU}{\affiliation{Department of Physics, The~University~of~Hong~Kong, Pokfulam, Hong~Kong}}
\newcommand{\UH}{\affiliation{Department of Physics, University~of~Houston, Houston, Texas, USA}}
\newcommand{\Charles}{\affiliation{Charles~University, Faculty~of~Mathematics~and~Physics, Prague}}
\newcommand{\USTC}{\affiliation{University~of~Science~and~Technology~of~China, Hefei}}
\newcommand{\TempleUniversity}{\affiliation{Department~of~Physics, College~of~Science~and~Technology, Temple~University, Philadelphia, Pennsylvania, USA}}
\newcommand{\CUC}{\affiliation{Instituto de F\'isica, Pontificia Universidad Cat\'olica de Chile, Santiago, Chile}} % update 20140911 use r23215 for Spanish
\newcommand{\CGNPG}{\affiliation{China General Nuclear Power Group}}% updated 20140724 China~Guangdong~Nuclear~Power~Group, Shenzhen}}
\newcommand{\NUDT}{\affiliation{College of Electronic Science and Engineering, National University of Defense Technology, Changsha}} % added 20140111
\newcommand{\IowaState}{\affiliation{Iowa~State~University, Ames, Iowa, USA}}
\newcommand{\ZSU}{\affiliation{Sun Yat-Sen (Zhongshan) University, Guangzhou}}
\newcommand{\CQU}{\affiliation{Chongqing University, Chongqing}} % add 20150417
\author{F.~P.~An}\ECUST
\author{A.~B.~Balantekin}\Wisconsin
\author{H.~R.~Band}\Yale
\author{M.~Bishai}\BNL
\author{S.~Blyth}\NTU\NUU
\author{I.~Butorov}\Dubna
\author{D.~Cao}\NJU
\author{G.~F.~Cao}\IHEP
\author{J.~Cao}\IHEP
\author{W.~R.~Cen}\IHEP
% ...

%\author[]{Daya~Bay~Collaboration}

\collaboration{The Daya Bay Collaboration}\noaffiliation
\date{\today}

%%%%%% Abstract %%%%%%%%%%%%
%%%%%% %%%%%%%%%%%%%%%%
\begin{abstract}
This Letter reports a measurement of the flux and energy spectrum of electron 
antineutrinos from six 2.9~GW$_{th}$ nuclear reactors with six 
detectors deployed in two near (effective baselines 512~m and 561~m) and one 
far (1,579~m) underground experimental halls in the Daya Bay experiment. 
Using 217 days of data, 296,721 and 41,589 inverse beta decay (IBD) 
candidates were detected in the near and far halls, respectively. 
The measured IBD yield is (1.55 $\pm$ 0.04) $\times$ 10$^{-18}$~cm$^2$/GW/day or (5.92 $\pm$ 0.14) $\times$ 10$^{-43}$~cm$^2$/fission. 
This flux measurement is consistent with previous short-baseline reactor antineutrino experiments and is $0.946\pm0.022$ ($0.991\pm0.023$) relative to the flux predicted with the Huber+Mueller (ILL+Vogel) fissile antineutrino model. 
The measured IBD positron energy spectrum deviates from both spectral predictions by more than 2$\sigma$ over the 
full energy range with a local significance of up to $\sim$4$\sigma$ between 4-6 MeV. 
A reactor antineutrino spectrum of IBD reactions is extracted from the measured positron energy spectrum 
for model-independent predictions.
\end{abstract}

\maketitle

%%%%%% Introduction  %%%%%%%%%%%%%
%%%%%%%%%%%%%%%%%%%%%%%%%%
Reactor antineutrino experiments have played a key role in developing the picture of neutrinos in the Standard
Model of particle physics. 
They provided the first experimental
observation of (anti)neutrinos~\cite{cowan1956}, 
confirmed neutrino oscillation as the solution to the solar neutrino 
problem~\cite{KamLAND}, provided the first detection of geo-neutrinos~\cite{bib:geo_araki}, 
and discovered the non-zero neutrino mixing angle $\theta_{13}$~\cite{bib:prl_rate,bib:RENO}. 
Forthcoming reactor antineutrino experiments are aiming to further explore the nature of neutrinos
by determining the neutrino mass hierarchy, precisely measuring neutrino mixing parameters, and searching
for short-baseline neutrino oscillation~\cite{bib:reactorExp}. 
Over the last five decades, reactor antineutrino experiments have measured the flux and spectrum of antineutrinos at various distances from nuclear reactors ranging from $\sim$10~m to several hundred kilometers.  
These measurements were found to be in good agreement~\cite{reactor_review} with predictions derived from the measurements of the beta spectra at ILL~\cite{bib:ILL_1,bib:ILL_2,bib:ILL_3} and Vogel's theoretical calculation~\cite{bib:vogel} when considering the effect of three-neutrino oscillation.  
In 2011, re-evaluations of the reactor antineutrino flux and spectrum with
improved theoretical treatments were carried out~\cite{bib:mueller2011,bib:huber}, and 
determined the flux to be higher than the experimental data. 
This discrepancy is commonly referred to as the ``Reactor Antineutrino Anomaly''~\cite{bib:mention2011} and may be a sign of new physics or insufficient fissile antineutrino modeling. 
Precision measurements by modern reactor antineutrino experiments can shed light on this issue and probe 
the physics underlying current reactor antineutrino predictions.
An accurate determination of the reactor antineutrino 
spectrum can also provide valuable input to next-generation 
single-detector reactor antineutrino experiments~\cite{bib:junoyb}.

This Letter reports measurements of the reactor
antineutrino flux and spectrum based on 217 days of data
from the Daya Bay experiment. 
The Daya Bay reactor complex consists of three nuclear power plants (NPPs),
each hosting two pressurized-water reactors. The maximum thermal power
of each reactor is 2.9~GW$_{th}$. 
The data used for this analysis comprises 338,310 antineutrino
candidate events collected in
six antineutrino detectors (ADs) in the two near
experimental halls (effective baselines 512~m and 561~m) and the one far hall
(effective baseline 1,579~m). This is the largest sample of reactor antineutrinos, comparable to that from the BUGEY-4 experiment~\cite{bib:bugey4}. 
A more detailed description of the experimental setup and the data set 
is given in Ref.~\cite{bib:prl_shape}.

%%%%%% Flux Prediction %%%%%%%%%%%%%
%%%%%%%%%%%%%%%%%%%%%%%%%%%
In reactor cores, electron antineutrinos ($\bar{\nu}_{e}$) are emitted isotropically from fission products of four primary isotopes: $^{235}$U, $^{238}$U, $^{239}$Pu, and $^{241}$Pu. The number of $\bar{\nu}_{e}$ with energy $E$ emitted from a reactor at a time $t$ can be predicted using
\begin{equation}\label{equ_antinu_prod}
\resizebox{\hsize}{!}{$\displaystyle\frac{d^{2}\phi(E, t)}{dE dt} = \frac{W_{th}(t)}{\sum_{i}f_{i}(t)e_{i}}\sum_{i}f_{i}(t)S_{i}(E)c^{ne}_{i}(E, t) + S_\textrm{SNF}(E, t),$}
\end{equation}
where the sums are over the four primary isotopes, $W_{th}(t)$ is the reactor thermal power, $f_{i}(t)$ is the fraction of fissions due to isotope $i$, $e_{i}$ is the average thermal energy released per fission, $S_{i}(E)$ is the $\bar{\nu}_{e}$ energy spectrum per fission, $c^{ne}_{i}(E, t)$ is the correction to the energy spectrum due to reactor non-equilibrium effects of long-lived fission fragments, and $S_\textrm{SNF}(E, t)$ is the contribution from spent nuclear fuel (SNF). At Daya Bay, the NPPs monitor the reactor power in real-time 
and simulate the evolution of the fuel composition using the SCIENCE software package~\cite{bib:science2004, bib:science2010}. The measured power (0.5\% uncertainty~\cite{bib:caojPower, bib:tournu, bib:npp2}) and simulated fission fractions ($\sim$5\% relative uncertainty~\cite{bib:apollo}) of each core are provided to the Daya Bay collaboration. Simulation of reactor cores based on DRAGON~\cite{bib:dragon} was constructed to study the correlations among the fission fractions of the four isotopes~\cite{bib:fissFrac}. The energies released per fission (0.2-0.5\% uncertainty) were from Ref.~\cite{bib:kopeikin}. Non-equilibrium (30\% uncertainty) and SNF (100\% uncertainty) corrections were applied following Refs.~\cite{bib:mueller2011} and \cite{bib:fengpeng, bib:zhoubin}, with $\sim$0.5\% and $\sim$0.3\% contributions to the total antineutrino rate, respectively. 
Combining the uncertainties of reactor power, fission fractions, and non-equilibrium and SNF corrections, the total reactor-uncorrelated uncertainty of antineutrino flux is 0.9\%.

Two fissile antineutrino spectrum models were used for $S_{i}(E)$ in Eq.~(\ref{equ_antinu_prod}) to predict the reactor antineutrino flux and spectrum. The ILL+Vogel model refers to the conventional ILL model~\cite{bib:ILL_1,bib:ILL_2,bib:ILL_3} of $^{235}$U, $^{239}$Pu, and $^{241}$Pu, and the theoretical model of $^{238}$U from Vogel~\cite{bib:vogel}.  The Huber+Mueller model refers to the recent re-evaluation of $^{235}$U, $^{239}$Pu and $^{241}$Pu from Huber~\cite{bib:huber}, and that of $^{238}$U from Mueller et al.~\cite{bib:mueller2011}. The Huber+Mueller model was chosen as a reference because of its improved theoretical treatments in beta-to-antineutrino conversions, and the information it provides about uncertainties and their correlations.  

Reactor antineutrinos were detected via inverse beta decay (IBD) reactions in the gadolinium-doped liquid scintillator (GdLS) of the Daya Bay ADs. The total number of detected IBD events $T$ in a given AD was estimated as
\begin{equation}\label{equ_ibd_rate}
\resizebox{\hsize}{!}{$\displaystyle T = \sum_{i=1}^6 \frac{N_{\!P}\  \varepsilon_{\mathrm{IBD}}}{4\pi L_i^2} \! \iint
\!P_{\mathrm{sur}}(E,L_i) \sigma_{\mathrm{IBD}}(E) 
\frac{d^{2}\phi_i(E, t)}{dE dt} dE dt,$}
\end{equation}
where $d^{2}\phi_i(E, t)/dE dt$ is the differential antineutrino rate from the $i$-th reactor core given in Eq.~(\ref{equ_antinu_prod}), 
$\sigma_{\mathrm{IBD}}(E)$ is the cross section of the IBD reaction, 
$L_i$ is the distance between the center of the detector and the $i$-th core~\cite{bib:cpc_rate},  
$P_{\mathrm{sur}}(E,L_i)$ is the survival probability due to neutrino oscillation, 
$N_{\!P}$ is the number of target protons~\footnote{Values are supplied with the supplemental material.}, and
$\varepsilon_{\mathrm{IBD}}$ is the efficiency of detecting IBD reactions.
The cross section $\sigma_{\mathrm{IBD}}(E)$ was evaluated based
on the formalism in Ref.~\cite{Vogel:1999zy}.
Physical constants including the neutron lifetime
(880.3 $\pm$ 1.1~s) were taken from the Particle Data Group~\cite{Agashe:2014kda}.

%%%%%% IBD Detection %%%%%%%%%%%%%%%
%%%%%%%%%%%%%%%%%%%%%%%%%%%%%
IBD candidates were selected by requiring a time coincidence between a prompt signal from an IBD positron including its annihilation energy, 
and a delayed signal from an IBD neutron after capturing on Gd, as described in Refs.~\cite{bib:prl_shape,bib:prl_rate}.  
The energy of interacting antineutrinos, $E$, is closely related to the prompt energy of the IBD positrons, $E_\text{prompt}$: without detector effects, $E_\text{prompt} \simeq E + (M_p - M_n - M_e) + 2M_e = E - 0.78$~MeV, where $M_p$, $M_n$, and $M_e$ are the proton, neutron, and electron masses. 
The reported dataset includes 296,721 and 41,589 IBD candidates at the near and far halls, respectively.
Corresponding background rates and spectra were estimated in Ref.~\cite{bib:prl_shape}, with about 5,470 $\pm$ 240 and 1,894 $\pm$ 43 background events at the near and far halls, respectively.

The relative analysis of IBD rates for $\sin^22\theta_{13}$ requires estimates of uncertainties that are uncorrelated among ADs while the measurement of flux is dominated by uncertainties that are correlated among ADs.  A detailed study of the event selection efficiencies was carried out using Monte Carlo simulation (MC)-data comparisons with the Daya Bay simulation framework based on \textsc{Geant4}~\cite{bib:Geant4}.  The previous study of efficiencies is described in detail in Ref.~\cite{bib:cpc_rate}.  
Estimates of each selection efficiency and detector characteristic are summarized in Table~\ref{tab:eff} and are briefly described in order below. 
Efficiencies of flashing-photomultiplier tube (flasher), capture-time, and prompt-energy selections were determined as described in Ref.~\cite{bib:cpc_rate} utilizing an updated Daya Bay IBD MC.
The IBD neutron-Gd capture fraction is dependent on the target's Gd concentration and on the escape, or `spill-out', of IBD neutrons from the target.
The former has been measured using neutron calibration sources deployed at the detector center, while the latter was estimated with MC-data comparisons of source deployments throughout the GdLS volume with a manual calibration system~\cite{bib:mcs}.
The efficiency for detecting Gd-capture IBD neutrons, also called the delayed energy cut efficiency, is dependent on the amount of Gd-capture $\gamma$ energy deposited outside the scintillator, and was determined using MC benchmarked to the IBD Gd-capture spectrum from data.
Finally, in order to account for contributions from IBD interactions outside the GdLS target, we have applied a spill-in correction determined using MC-data comparisons of IBD coincidence time and reconstructed position distributions.
The updated detector efficiency $\varepsilon$ was estimated to be 80.6\% with an AD-correlated fractional uncertainty $\delta_{\varepsilon}$/$\varepsilon$ of 2.1\%.  
Application of additional AD-dependent muon veto and multiplicity cut efficiencies, described in detail in Refs.~\cite{bib:cpc_rate,bib:prl_shape}, produced total detection efficiencies $\varepsilon_{\mathrm{IBD}}$ ranging from 64.6\% to 77.2\% among ADs. 
The total correlated uncertainty was dominated by the spill-in correction, 
whose uncertainty enveloped the individual uncertainties provided by 
three independent methods and was limited by small biases in position reconstruction. 
A cross-check of the spill-in effect provided by data-MC comparisons of neutron sources deployed outside the target volume showed agreement well within this uncertainty.

\begin{table}[tbp]
\caption{Summary of IBD selection efficiencies and their AD-correlated uncertainties. The uncertainties are given in relative units. } 
\centering \begin{tabular}{l | c | c}
\hline
\hline
                           &  Efficiency ($\varepsilon$)      &   Uncertainty ($\delta_{\varepsilon}$/$\varepsilon$)   \\
 \hline
 Target protons         &   -                      & 0.47\% \\
Flasher cut                & 99.98\%         & 0.01\%    \\
Capture-time cut      &    98.70\%       &  0.12\%   \\
Prompt-energy cut  & 99.81\%           &  0.10\%   \\
Gd-capture fraction       &     84.17\%       &   0.95\% \\
Delayed-energy cut     &    92.71\%          &  0.97\%    \\
Spill-in correction      &     104.86\%       & 1.50\%   \\
\hline
Combined                   &   80.6\%           &   2.1\%   \\
\hline
\hline
\end{tabular}
\label{tab:eff}
\end{table} 

%%%%%% Flux Normalization %%%%%%%%%%%%%%
%%%%%%%%%%%%%%%%%%%%%%%%%%%%%%
To extract the rate of IBD interactions at Daya Bay, the $\theta_{13}$-driven oscillation effect must be corrected for in each detector. 
A normalization factor $R$ was defined to scale the measured rate to that predicted with a fissile antineutrino spectrum model. The value of $R$, together with the value of $\sin^22\theta_{13}$, were simultaneously determined with a $\chi^2$ similar to the one used in Ref.~\cite{bib:prl_rate}: 
\begin{multline}\label{equ_norm}
\chi^{2} = \sum_{d=1}^{6} \frac{[M_{d} - R \cdot T_{d}(1 + \epsilon_{D} + \sum_{r}\omega_{r}^{d}\alpha_{r} + \epsilon_{d}) + \eta_{d}]^{2}}{M_{d} + B_{d}} \\
+ \sum_{r}\frac{\alpha_{r}^{2}}{\sigma_{r}^{2}} + \sum_{d=1}^{6}\biggl(\frac{\epsilon_{d}^{2}}{\sigma_{d}^{2}} + \frac{\eta_{d}^{2}}{\sigma_{B_d}^{2}}\biggr) + \frac{\epsilon_{D}^{2}}{\sigma_{D}^{2}},
\end{multline}
where $M_{d}$ is the number of measured IBD events in the $d$-th detector with backgrounds subtracted, $B_{d}$ is the corresponding number of background events, $T_{d}$ is the number of IBD events predicted with a fissile antineutrino spectrum model via Eq.~(\ref{equ_ibd_rate}), and $\omega_{r}^{d}$ is the fractional IBD contribution from the $r$-th reactor to the $d$-th detector determined with baselines and reactor antineutrino rates, $\sigma_{r}$ (0.9\%) is the uncorrelated reactor uncertainty, $\sigma_{d}$ (0.2\%~\cite{bib:prl_shape}) is the uncorrelated detection uncertainty, $\sigma_{B_d}$ is the background uncertainty listed in Ref.~\cite{bib:prl_shape}, and $\sigma_{D}$ (2.1\%) is the correlated detection uncertainty, i.e. the uncertainty of detection efficiency in Table~\ref{tab:eff}. Their corresponding nuisance parameters are $\alpha_{r}$, $\epsilon_{d}$, $\eta_{d}$, and $\epsilon_{D}$, respectively.
The best-fit value of $\sin^{2}2\theta_{13} = 0.090 \pm 0.009$ is insensitive to the choice of model. The best-fit value of $R$ is $0.946 \pm 0.022$ ($0.991 \pm 0.023$) when predicting with the Huber+Mueller (ILL+Vogel) model. Replacing the Mueller $^{238}$U spectrum with the recently-measured spectrum in Ref.~\cite{bib:munich} yields negligible change in $R$. The uncertainty in $R$ is dominated by the correlated detection uncertainty $\sigma_{D}$. 

\begin{figure}[bp] 
\centering
\includegraphics[width=\columnwidth]{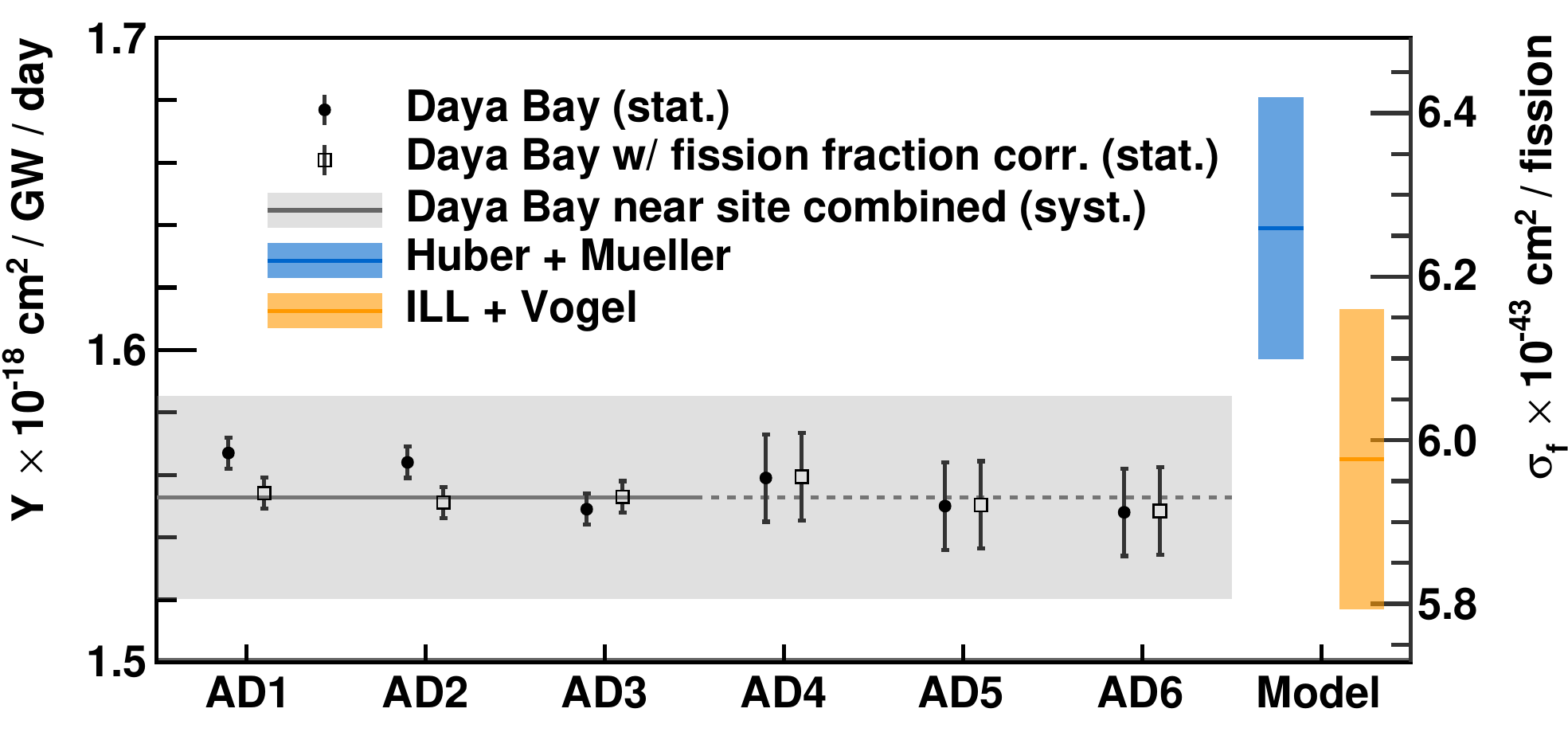} 
\includegraphics[width=\columnwidth]{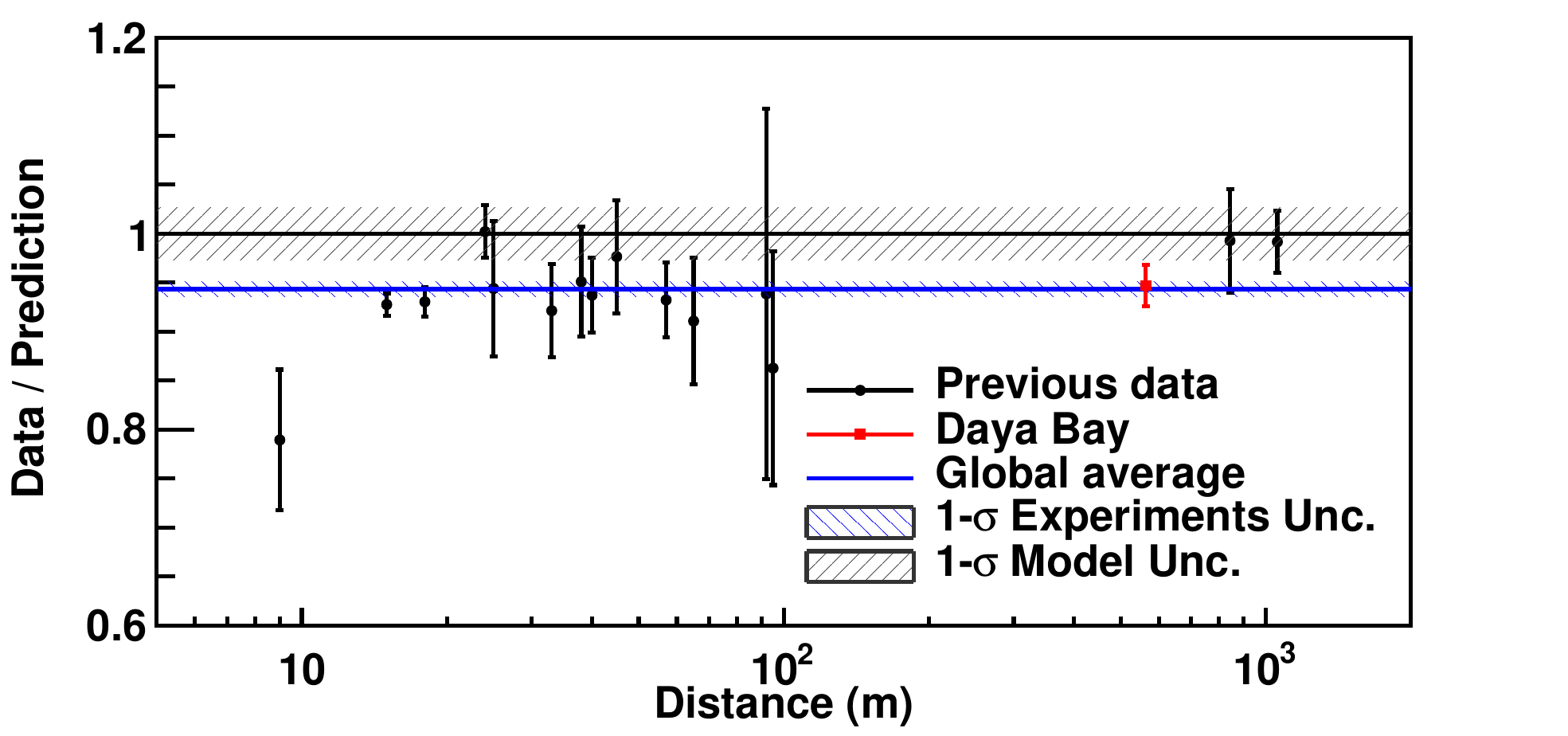} 
\caption{Top: Rate of reactor antineutrino candidate events in the six ADs with corrections for 3-flavor oscillations (closed circles), and additionally for the variation of flux-weighted fission fractions at the different sites (open squares). The average of the three near detectors is shown as a gray line (and extended through the three far detectors as a dotted gray line) with its $1\sigma$ systematic uncertainty (gray band). The rate predicted with the Huber+Mueller (ILL+Vogel) model and its uncertainty are shown in blue (orange). Bottom: The measured reactor $\bar{\nu}_e$ rate as a function of the distance from the reactor, normalized to the theoretical prediction with the Huber+Mueller model. The rate is corrected for 3-flavor neutrino oscillations at each baseline. The blue shaded region represents the global average and its $1\sigma$ uncertainty. The 2.7\% model uncertainty is shown as a band around unity. Measurements at the same baseline are combined for clarity. The Daya Bay measurement is shown at the flux-weighted baseline (573 m) of the two near halls.}
\label{fig:norm}
\end{figure}

With the oscillation effect for each AD corrected using the best-fit value of $\sin^22\theta_{13}$ in Eq.~(\ref{equ_norm}), the measured IBD yield for each AD is expressed in two ways: the yield per GW$_{th}$ per day, $Y$, and equivalently, the yield per nuclear fission, $\sigma_f$. These results are shown in the top panel of Fig.~\ref{fig:norm}. 
The measured IBD yields are consistent among all ADs after further correcting for the small variations of fission fractions among the different sites. The average IBD yield in the three near ADs is $Y = (1.55 \pm 0.04) \times 10^{-18}$~cm$^2$/GW/day, or $\sigma_f = (5.92 \pm 0.14) \times 10^{-43}$~cm$^2$/fission. These results are summarized in Table~\ref{tab:norm3m} along with the flux-weighted average fission fractions in the three near ADs.

\begin{table}[tbp]
\caption{Average IBD yields ($Y$ and $\sigma_f$) of the near halls, flux normalization with respect to different fissile antineutrino model predictions, and flux-weighted average fission fractions of the near halls.} 
\centering \begin{tabular}{c | c}
\hline
\hline
\multicolumn{2}{c}{IBD Yield} \\
\hline
$Y$~(~cm$^2$/GW/day)       &  $(1.55 \pm 0.04) \times 10^{-18}$          \\
$\sigma_f$~(cm$^2$/fission)     &  $(5.92 \pm 0.14) \times 10^{-43}$          \\
\hline
\multicolumn{2}{c}{Data / Prediction} \\
\hline
$R$~(Huber+Mueller) &  $0.946 \pm 0.022$ \\
$R$~(ILL+Vogel) & $0.991 \pm 0.023$ \\
\hline
$^{235}$U : $^{238}$U : $^{239}$Pu : $^{241}$Pu    ~       &  ~  0.586 : 0.076 : 0.288 : 0.050 \\
\hline
\hline
\end{tabular}
\label{tab:norm3m}
\end{table} 

A global fit for $R$ was performed to compare with previous reactor antineutrino flux measurements following the method described in Ref.~\cite{bib:chao}. Nineteen past short-baseline ($<$100~m) measurements were included using the data from Ref.~\cite{bib:mention2011}. The measurements from CHOOZ~\cite{bib:chooz} and Palo Verde~\cite{bib:paloverde} were also included after correcting for the effect of standard three-neutrino oscillations. All measurements were compared to the Huber+Mueller model. All predictions were fixed at their nominal value in the fit. The resulting past global average is $R_g^{past} = 0.942 \pm 0.009 \, \textrm{(exp.)} \pm 0.025 \, \textrm{~(model)}$.  Daya Bay's measurement of the reactor antineutrino flux is consistent with the past experiments. 
Including Daya Bay in the global fit, the new average is $R_g = 0.943 \pm 0.008 \, \textrm{(exp.)} \pm 0.025 \, \textrm{(model)}$.  
The results of the global fit are shown in the bottom panel of Fig.~\ref{fig:norm}.

%%%% Spectrum Comparison %%%%%%%%%%%%%
%%%% %%%%%%%%%%%%%%%%%%%%%%%%%
Extending the study from the integrated flux to the energy spectrum, 
the measured prompt-energy spectra of the three near-site ADs were combined 
after background subtraction and compared with predictions. 
The antineutrino spectrum at each detector was predicted by the
procedure described above, taking into account neutrino oscillation with 
$\sin^22\theta_{13} = 0.090$ and $\Delta m^2_{ee} = 2.59\times
10^{-3}~\mathrm{eV}^2$ based on the oscillation analysis of the same data~\cite{bib:prl_shape}.
The detector response was determined in two ways. 
The first method sequentially applied a simulation of energy
loss in the inactive acrylic vessels, and analytical models of energy
scale and energy resolution.
The energy scale model was 
based on empirical characterization of the spatial non-uniformity and
the energy non-linearity with improved calibration of the scintillator light yield and
the electronics response~\cite{bib:8AD_shape}.
The uncertainty of the energy scale was about 1\% in the energy range of
reactor antineutrinos~\cite{bib:8AD_shape}.
The second method used full-detector
simulation in which the detector response was tuned with the
calibration data.
Both methods produced consistent predictions 
for prompt energies above 1.25~MeV.
Around 1~MeV, there was a slight discrepancy due to different
treatments of IBD positrons that interact with the inner acrylic vessels. 
Additional uncertainty below 1.25~MeV was included to cover this discrepancy.  

\begin{figure}[htbp] 
\centering
\includegraphics[width=\columnwidth]{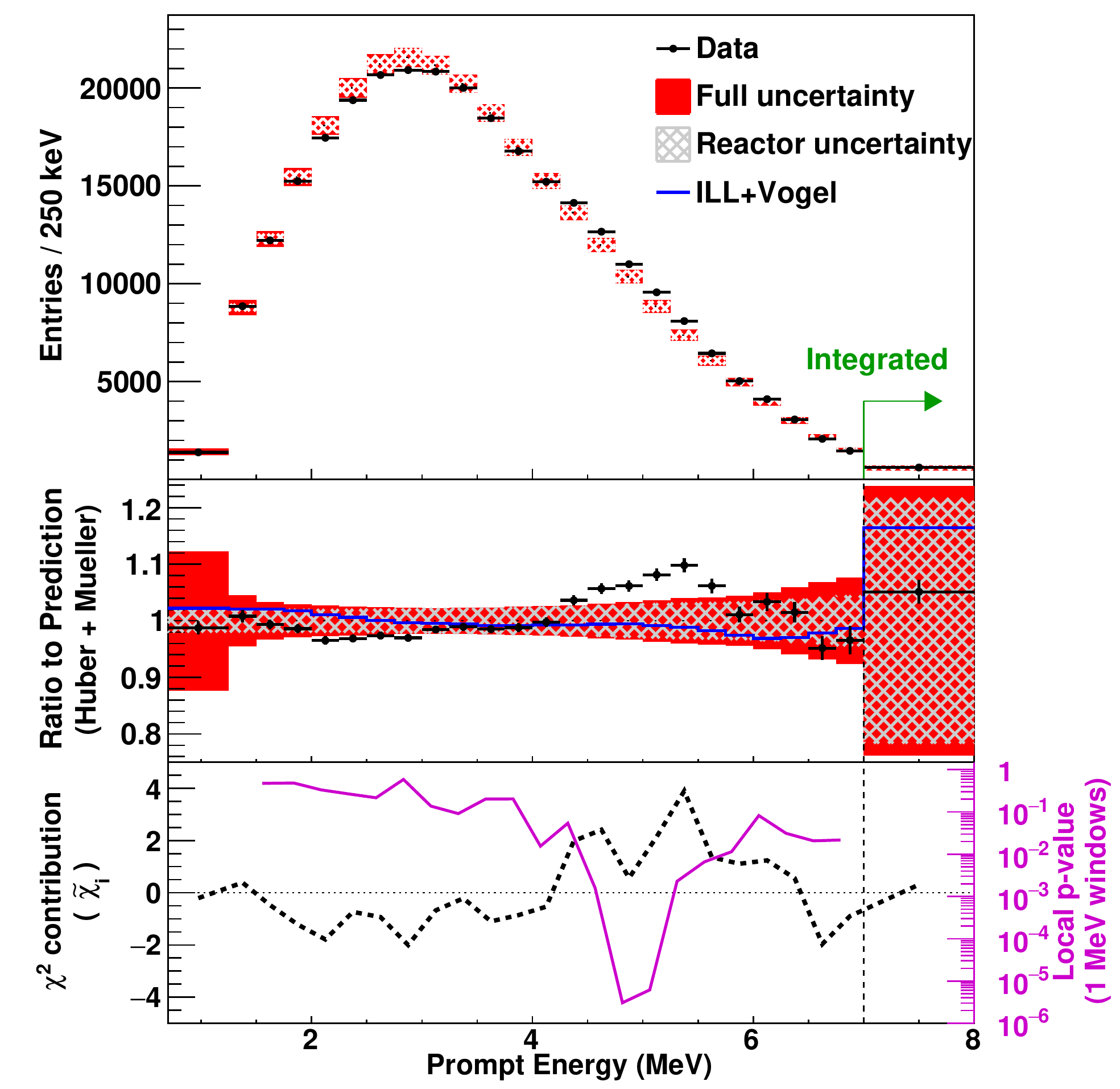} 
\caption{
  Top panel: Predicted and measured prompt-energy spectra.
  The prediction is based on the Huber+Mueller model and 
  normalized to the number of measured events. 
  The highest energy bin contains all events above 7~MeV.
  The gray hatched and red filled bands 
  represent the square-root of diagonal elements of 
  the covariance matrix ($\sqrt{V_{ii}}$) for the reactor related and
  the full (reactor, detector and background) systematic uncertainties, respectively.
  The error bars on the data points represent the statistical uncertainty.
  Middle panel: 
  Ratio of the measured prompt-energy spectrum to the predicted spectrum
  (Huber+Mueller model). 
  The blue curve shows the ratio of the prediction based
  on the ILL+Vogel model to that based on the Huber+Mueller model.
  Bottom panel: The defined $\chi^2$ distribution ($\widetilde{\chi_i}$) of 
  each bin (black dashed curve) and local p-values for 1-MeV energy
  windows (magenta solid curve). See the text
  for the definitions of these quantities.}
\label{fig:prompt}
\end{figure}

Figure~\ref{fig:prompt} shows the observed prompt-energy spectrum and its comparison with the predictions.
The spectral uncertainty of the measurement is 
composed of the statistical, detector response and background uncertainties. Between 1.5 and 7 MeV, it ranges from 1.0\% at 3.5 MeV
to 6.7\% at 7 MeV, and above 7 MeV it is larger than 10\%. 
The predicted spectra were normalized to the measurement thus removing the dependence on the total rate. Agreement between a prediction and the data was quantified with the $\chi^2$ defined as
\begin{equation}
  \label{eq:chi2_prompt}
  \chi^2 = \sum_{i,j}(N^{\mathrm{obs}}_i - N^{\mathrm{pred}}_i)V^{-1}_{ij}(N^{\mathrm{obs}}_j - N^{\mathrm{pred}}_j),
\end{equation}
where $N^{\mathrm{obs(pred)}}_i$ is the observed (predicted) number of events at the
$i$-th prompt-energy bin and $V$ is the covariance matrix that includes all 
statistical and systematic uncertainties. 
The systematic uncertainty portion of the covariance matrix $V$ was
estimated using simulated data sets with randomly fluctuated detector response, background
contributions, and reactor-related uncertainties, while the statistical uncertainty portion
was calculated analytically. A comparison to the Huber+Mueller model yielded a
$\chi^2$/NDF, where NDF is the number of
degrees of freedom, of 43.2/24 in the full
energy range from 0.7 to 12~MeV, corresponding to a 2.6$\sigma$
discrepancy.
The ILL+Vogel model showed a similar level of discrepancy from the data at 2.4$\sigma$.  

%%%%% Local quantification of the discrepancy between Data and Prediction %%%
%%%%%%%%%%%%%%%%%%%%%%%%%%%%%%%%%%%%%%%%%%
The ratio of the measured to predicted prompt-energy spectra is shown 
in the middle panel of Fig.~\ref{fig:prompt}. A discrepancy is apparent around 5~MeV. 
Two approaches were adopted to evaluate the significance of local discrepancies. 
The first was based on the $\chi^{2}$ contribution of each energy bin, 
which is evaluated by 
\begin{align}
\begin{split}
\widetilde{\chi_i} & = \frac{N^{\rm obs}_i - N^{\rm pred}_i}{\left|N^{\rm obs}_i - N^{\rm pred}_i\right|} \sqrt {\sum_{j}\chi^2_{ij}}, \\ 
\text{where} ~& \chi^2_{ij} \equiv (N^{\rm obs}_i - N^{\rm pred}_i)V^{-1}_{ij}(N^{\rm obs}_j - N^{\rm pred}_j).
\end{split}
\end{align}
As shown in the bottom panel of Fig.~\ref{fig:prompt}, there is a larger
contribution around 5~MeV. 
In the second approach, the significance of deviations are conveyed with 
p-values calculated within local energy windows. 
A free-floating nuisance parameter for the normalization of each bin
within a chosen energy window was introduced to the fitter that was used in the neutrino oscillation analysis. 
The difference in the minimum $\chi^2$ before and after introducing 
these nuisance parameters was used to evaluate the p-value of the 
deviation from the theoretical prediction within each window.
The p-values within 1-MeV energy windows are shown in the bottom panel of 
Fig.~\ref{fig:prompt}. The p-value for a 2-MeV window between 
4 and 6~MeV reached a similar minimum of $5.4\times10^{-5}$, which corresponds
to a $4.0\sigma$ deviation. The ILL+Vogel model showed a similar level of discrepancy 
between 4 and 6~MeV. 

The number of events in excess of the predictions in the 4-6~MeV region was estimated to comprise approximately 1\% of all events in both the near and far detectors. This excess is approximately 10\% of events within the 4-6~MeV region. 
This discrepancy was found to be time-independent and correlated 
with reactor power, therefore disfavoring 
hypotheses involving detector response and unknown backgrounds.
A recent ab-initio calculation of the antineutrino spectrum
showed a similar deviation from previous predictions in the 4-6~MeV region,
and identified prominent fission daughter isotopes as a potential 
explanation~\cite{bib:dan}. 
A number of tentative explanations based on the nuclear physics of beta decays and fission yields
have been put forward and are under active investigation; for examples, see Refs.~\cite{bib:dan, bib:sonzogni, bib:hayes, bib:zakari, bib:hayes2}. These studies suggest an increased uncertainty in both the yields and spectra of the fissile antineutrino models, which may also account for the discrepancy. 

%%%%%% Generic antineutrino spectrum %%%%%%%%%%%%%%
%%%%%% %%%%%%%%%%%%%%%%%%%%%%%%%%%%%
From the measured IBD prompt spectrum at Daya Bay, we have obtained a reactor antineutrino spectrum of IBD reactions that can be used to make model-independent predictions of reactor antineutrino flux and spectra~\footnote{An example is supplied with the supplemental material.}.  
The spectrum was obtained by first summing the prompt-energy spectra of the three near site ADs weighted with their target mass relative to the average target mass of all near-site ADs, $\overline{M}$: $S_\textrm{combined}(E_\text{prompt})$ = $\sum_{i=1}^3 S_i(E_\text{prompt})\overline{M}/M_i$. Detector response effects were then removed by unfolding the combined prompt spectrum $S_\textrm{combined}(E_\text{prompt})$ to an antineutrino spectrum of IBD reactions, $S_\textrm{combined}(E)$. Finally, oscillation effects were removed and each bin of the antineutrino spectrum was normalized to cm$^2$/fission/MeV using the thermal power $W_{th}(t)$ and fission fraction $f_{i}(t)$ information of each core. The reactor antineutrino spectrum is expressed as 
\begin{equation}\label{equ_generic}
S_\textrm{reactor} (E) = \frac{S_\textrm{combined}(E)} {\overline{P}_\textrm{sur}(E) \cdot \overline{N}_{\!P} \cdot F_\textrm{total}},
\end{equation}
where $\overline{P}_\textrm{sur}(E)$ is the 
flux-weighted average of the survival probabilities $P_{\mathrm{sur}}(E,L_{i,d})$ from the six reactors ($i$) to the three detectors ($d$), 
$\overline{N}_{\!P}$ is the number of target protons in $\overline{M}$, and $F_\textrm{total}$ is the total number of fissions 
from the sum of the fissions of the six reactors to the three detectors weighted with $\varepsilon_{\mathrm{IBD},d}/4\pi L^2_{i,d}$.  
Correcting the unfolded spectrum with an average survival probability resulted in a negligible bias ($<$0.01\%).  

\begin{figure}[t] 
\centering
\includegraphics[width=0.95\columnwidth]{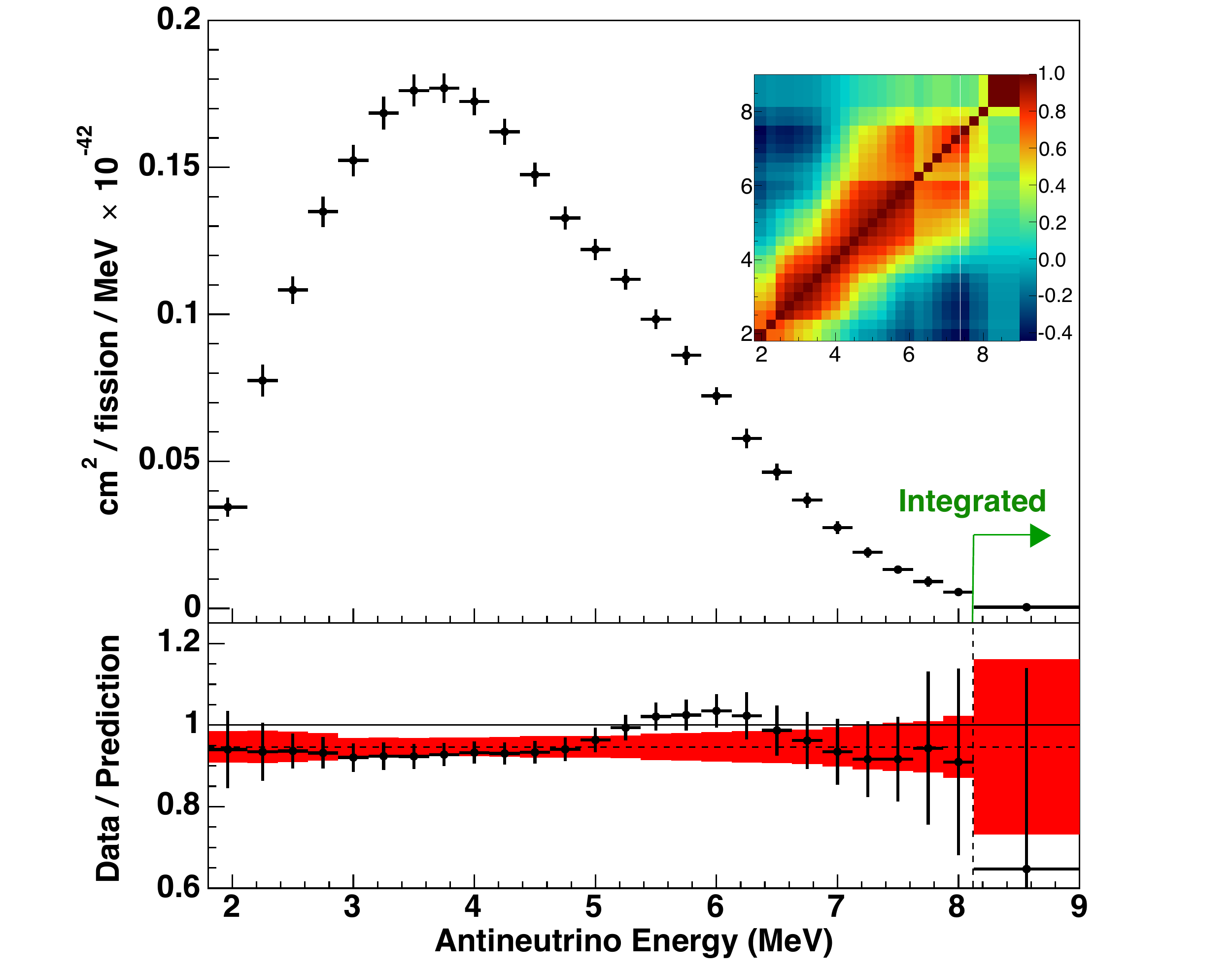} 
\caption{Top panel: The extracted reactor antineutrino spectrum and its correlation matrix. 
Bottom panel: Ratio of the extracted reactor antineutrino spectrum to the Huber+Mueller prediction. 
The error bars of the data points are the square-roots of the diagonal elements of the data covariance matrix, which included statistical and systematic uncertainties, as well as bias and the AD-correlated uncertainty from Table~\ref{tab:eff}. The solid red band represents the square-roots of the diagonal elements of the prediction covariance matrix, which included reactor and Huber+Mueller model uncertainties. The horizontal dashed line represents the normalization factor $R$ = 0.946. The vertical dashed line denotes that above 8~MeV, the Huber+Mueller model was extrapolated.}  
\label{fig:generic}
\end{figure}

Detector response effects were removed with the Singular Value Decomposition~(SVD) unfolding method~\cite{bib:svd}. Statistical and systematical uncertainties are naturally propagated in the SVD method. The bias of unfolding was estimated by using detector response matrices constructed from the two different detector response models and by using a variety of input antineutrino spectra which covered the uncertainties of the two models and those estimated in Ref.~\cite{bib:dan}. The bin-to-bin bias between 2.2 and 6.5~MeV was about 0.5\%, which was a few times smaller than the statistical uncertainty. The bias outside this region was about 4\% and increased with energy due to the decrease of events. The bias values were assigned as additional uncertainties to the unfolded spectrum. 
Unfolding performed with the Bayesian iteration method~\cite{bib:bayes, bib:rooUnfold} produced consistent results.  
Between 2 and 7.5 MeV, the spectral uncertainty of the unfolded spectrum $S_\textrm{combined}(E)$ ranges from 1.1\% at 4.25 MeV to 9.3\% at 7.5 MeV, and around 10 MeV is more than 20\% due to low statistics. 
The obtained reactor antineutrino spectrum and its correlation matrix are shown in the top panel of Fig.~\ref{fig:generic}. Between 2 and 7.5 MeV, the uncertainty of the diagonal elements ranges from 2.7\% at 4.25 MeV to 10.4\% at 7.5 MeV. The bottom panel of Fig.~\ref{fig:generic} is the ratio of the extracted reactor antineutrino spectrum to the prediction using the fissile antineutrino spectra of the Huber+Mueller model and the average fission fractions listed in Table~\ref{tab:norm3m}. The integral of the ratio is equal to the flux normalization factor $R$ given in Table~\ref{tab:norm3m}. The integral of the spectrum is equal to the yield $\sigma_f$ given in Table~\ref{tab:norm3m}. The discrepancy between 5 and 7~MeV corresponds to the discrepancy between 4 and 6~MeV in the IBD prompt-energy distribution in Fig.~\ref{fig:prompt}. 

%%%%% SUMMARY %%%%%%%%%%%%
%%%%%%%%%%%%%%%%%%%%%%%
In summary, the Daya Bay experiment collected more than 330,000 antineutrino events in the data-taking period with six antineutrino detectors. 
The measured IBD yield is (1.55 $\pm$ 0.04) $\times$ 10$^{-18}$~cm$^2$/GW/day or (5.92 $\pm$ 0.14) $\times$ 10$^{-43}$~cm$^2$/fission. 
This flux measurement is consistent with the global average of previous short baseline experiments and is $0.946\pm0.022$ $(0.991\pm0.023)$ times the prediction using the Huber+Mueller (ILL+Vogel) fissile antineutrino model. 
In addition, the measured and predicted spectra are discrepant with a significance of $\sim$4$\sigma$ in the 4-6 (5-7)~MeV region of the IBD prompt (antineutrino) energy spectrum. Investigation of the discrepancy strongly disfavors explanations involving detector response or an unknown background. 
A reactor antineutrino spectrum was extracted from the measurement at Daya Bay, enabling model-independent predictions of reactor antineutrino spectra. 

%%%%%% Acknowledgements %%%%%%%%
Daya Bay is supported in part by the Ministry of Science and Technology of China, the U.S. Department of Energy, the Chinese Academy of Sciences (CAS), the CAS Center for Excellence in Particle Physics, the National Natural Science Foundation of China, the Guangdong provincial government, the Shenzhen municipal government, the China General Nuclear Power Group, 
Laboratory Directed Research $\&$ Development Program of Institute of High Energy Physics,
Shanghai Laboratory for Particle Physics and Cosmology, the Research Grants Council of the Hong Kong Special Administrative Region of China, the University Development Fund of The University of Hong Kong, the MOE program for Research of Excellence at National Taiwan University, National Chiao-Tung University, and NSC fund support from Taiwan, the U.S. National Science Foundation, the Alfred~P.~Sloan Foundation, Laboratory Directed Research $\&$ Development Program of Berkeley National Laboratory and Brookhaven National Laboratory, the Ministry of Education, Youth, and Sports of the Czech Republic, Charles University in Prague, the Joint Institute of Nuclear Research in Dubna, Russia, the NSFC-RFBR joint research program, the National Commission of Scientific and Technological Research of Chile. 
We acknowledge Yellow River Engineering Consulting Co., Ltd., and China Railway 15th Bureau Group Co., Ltd., for building the underground laboratory. We are grateful for the ongoing cooperation from the China General Nuclear Power Group and China Light and Power Company.

%%%%%% References %%%%%%%%%%%%%
\bibliographystyle{apsrev4-1}
\bibliography{Reactor_6AD_PRL}{}

\end{document}